\documentclass[aps,prd,a4,onecolumn,superscriptaddress,nofootinbib,preprintnumbers,11pt]{revtex4}
\usepackage{amsfonts}
\usepackage{mathrsfs}
\usepackage{amssymb}
\usepackage{amsmath}
\usepackage{graphicx,epsfig,epstopdf}
\usepackage{pstricks}
\usepackage{slashed}
\usepackage{dcolumn}% \usepackage{graphicx,bm}

\usepackage{color}
\usepackage{slashed}
\usepackage{color}
\usepackage{appendix}
\usepackage{resizegather}
\usepackage{bm}
\usepackage{subfig}
\definecolor{nicered}{rgb}{0.7,0.1,0.1}
\definecolor{nicegreen}{rgb}{0.1,0.5,0.1}

\usepackage{array}
\newcolumntype{L}[1]{>{\raggedright\let\newline\\\arraybackslash\hspace{0pt}}m{#1}}
\newcolumntype{C}[1]{>{\centering\let\newline\\\arraybackslash\hspace{0pt}}m{#1}}
\newcolumntype{R}[1]{>{\raggedleft\let\newline\\\arraybackslash\hspace{0pt}}m{#1}}

\usepackage{setspace}
\usepackage{multirow}

\def\wt{\widetilde}
\def\({\left(}
\def\){\right)}
\def\[{\left[}
\def\]{\right]}

\newcommand{\ba}{\begin{array}}
\newcommand{\ea}{\end{array}}
\newcommand{\bd}{\begin{displaymath}}
\newcommand{\ed}{\end{displaymath}}
\newcommand{\be}{\begin{equation}}
\makeatletter
\newcommand*{\rom}[1]{\expandafter\@slowromancap\romannumeral #1@}
\makeatother
\newcommand{\ee}{\end{equation}}
\def\bt{\begin{table}}
\def\et{\end{table}}
\def\bc{\begin{center}}
\def\ec{\end{center}}
\def\bi{\begin{itemize}}
\def\ei{\end{itemize}}
\def\bw{\begin{widetext}}
\def\ew{\end{widetext}}

\def\bea{\begin{eqnarray}}
\def\eea{\end{eqnarray}}
\def\beas{\begin{eqnarray*}}
\def\eeas{\end{eqnarray*}}

\def\e{\ell}

\def\N0{\widetilde{\chi}^0}

\def\a{\alpha}

\def\d{\delta}
\def\e{\epsilon}

\def\ov {\overline}

\def\l{\lambda}
\begin{document}
\preprint{HRI-RECAPP-2017-014} 
\title{A Lepton-specific Universal Seesaw Model with Left-Right Symmetry}

\author{Ayon Patra} 
\email{ayon@okstate.edu}
\affiliation{Centre for High Energy Physics, Indian Institute of Science, Bangalore - 560012, India}

\author{Santosh Kumar Rai} 
\email{skrai@hri.res.in} 
\affiliation{Regional Centre for Accelerator-based Particle Physics,\\
Harish-Chandra Research Institute, HBNI, Jhusi, Allahabad - 211019, India}

%\pacs{}

\vskip 0.3cm 
\begin{abstract}
%%%%%%%%%%%%%%%%%%%%%%%%%%%%%%%%%%%%%%%%%%%%%%%%%%%%%%%
We propose a left-right symmetric framework with universal seesaw mechanism for the generation of masses of 
the Standard Model quarks and leptons. Heavy vector-like singlet quarks and leptons are required for generation 
of Standard Model-like quark and lepton masses through seesaw mechanism. A softly broken $Z_2$ symmetry 
distinguishes the lepton sector and the quark sector of the model. This leads to the presence of 
some lepton-specific interactions that can produce unique collider signatures which can be 
explored at the current Large Hadron Collider run and also future colliders.
\end{abstract}
%%%%%%%%%%%%%%%%%%%%%%%%%%%%%%%%%%%%%%%%%%%%%%%%%%%%%%%%
\maketitle
\section{Introduction}
\label{sec:intro}

Left-right symmetric (LRS) models \cite{lr} are one of the most well motivated and widely studied extensions of 
the Standard Model (SM). The popularity of LRS models stem from the fact that in these models it is possible 
to explain several phenomenon which are not very well understood in the framework of SM. Fundamentally 
parity (P) is a good symmetry in these models and can be spontaneously broken at some high scale leading 
to a SM-like gauge structure at the electroweak scale. Thus we can understand the origin of parity violation as 
a spontaneously broken symmetry rather than it being explicitly broken. Parity symmetry also prevents one 
from writing P and Charge-Parity (CP) violating terms in the Quantum Chromodynamic (QCD) Lagrangian. 
Since CP violating terms in the color sector are highly constrained from neutron electric dipole measurements, 
the absence of these terms can solve the strong CP problem \cite{scp} naturally without the need to introduce a global 
Peccei-Quinn symmetry \cite{pq}. The gauge structure of these models forces us to have a right-handed neutrino 
in the lepton multiplet. This right-handed neutrino can generate a light neutrino mass through the 
seesaw mechanism \cite{csaw}.

Generally, in LRS models, an $SU(2)_R$ triplet Higgs boson is responsible for generation of the right-handed 
neutrino mass while a bidoublet field is needed to produce the quark and lepton masses and CKM mixings. All these multitude of scalar fields make the scalar sector quite complicated \footnote{A detailed study of various scalar sectors in LRS models is discussed in \cite{AB}}. It would be interesting, on the 
other hand, to consider a Higgs spectrum consisting purely of doublets. This would be similar to the Two 
Higgs Doublet model (2HDM) but we would need four doublets instead of two (two similar to 2HDM and other 
two their right-handed counterparts). The model we study here is a lepton-specific scenario where one pair 
of Higgs doublets couple only to the leptons. This has the distinct advantage that the quark and charged 
lepton masses can be generated keeping the Yukawa couplings to be of the same order for each generation. 
Thus we can easily avoid the large hierarchy observed in the Yukawa sector of the SM. 

To arrange the lepton-specific framework, we need to introduce an extra $Z_2$ symmetry under which a 
couple of Higgs boson doublets as well as the heavy lepton singlet fields are odd, all other fields being even. 
As these odd-$Z_2$ Higgs bosons get a non-zero vacuum expectation value (VEV), one expects this 
discrete $Z_2$ symmetry to be spontaneously broken. This could lead to domain walls and can make the model unstable from a cosmological point of view \cite{domain}. Such instabilities however can be avoided by introducing soft-breaking terms in the scalar 
potential. The consequence of these terms is that it leads to mixing between the different scalars in the 
doublets and can lead to interesting phenomenology. With the Large Hadron Collider (LHC) running, it 
is imperative to consider different scenarios for signals beyond the SM. In that spirit our 
model within the framework of left-right symmetry proposes new signals arising from a 
lepton-specific framework which generates all the SM fermion masses and gives lepton rich final 
states that could be observed or excluded at LHC.

All the fermion masses in this case are generated through universal seesaw 
mechanism \cite{UCsaw} by introducing singlet fermionic states. Most of the charged singlet fermions 
are quite heavy except the top quark partner to some extent, which is required to 
be lighter than the others and of the order of a few TeV. The other low lying states are the 
heavy neutral leptons and some extra scalars in the model.

The rest of the paper is organized as follows. In Section \ref{sec:model}, we discuss our model in detail 
and in Section \ref{sec:pheno} we discuss the phenomenological implications of our 
model including the experimental constraints and possible collider signatures. 
Section \ref{sec:concl} contains our conclusions and discussions.

\section{Model and Lagrangian}
\label{sec:model}

We consider a left-right symmetric model with the gauge group being 
$SU(3)_C \times SU(2)_L \times SU(2)_R \times U(1)_{B-L}$. An extra $Z_2$ symmetry is 
introduced which prevents several interactions facilitating a lepton-specific scenario. The 
charge of a particle in this model is defined as:
\begin{equation}
\mathcal{Q}=I_{3L}+I_{3R}+\frac{B-L}{2}.
\end{equation}
The chiral matter fields consist of three families of quark and lepton :
\begin{eqnarray}
\!\!Q_L\!\!&=&\!\!\left (\begin{array}{c}
u\\ d \end{array} \right )_L \sim \left (3,2, 1, \frac13 \right ),
Q_R\!=\!\left (\begin{array}{c}
u\\d \end{array} \right )_R \sim \left ( 3,1, 2, \frac13
\right ),\nonumber \\
l_L&=&\left (\begin{array}{c}
\nu\\ e\end{array}\right )_L \sim\left ( 1,2, 1, -1 \right ),
l_R=\left (\begin{array}{c}
\nu \\ e \end{array}\right )_R \sim \left ( 1,1, 2, -1 \right ),
\end{eqnarray}
where the numbers in the parentheses denote the quantum numbers under
$SU(3)_C \times SU(2)_L \times SU(2)_R \times U(1)_{B-L}$ gauge groups respectively. 

%\begin{widetext}
% 
\begin{table}
\begin{center}
{\renewcommand{\arraystretch}{1.5}
\begin{tabular}{c|c|c|c|c|c} \hline
Field & $SU(3)_C$ & $SU(2)_L$ & $SU(2)_R$ & $U(1)_{B-L}$ & ${Z}_2$ \\ \hline
$Q_L = \left (\begin{array}{c} u\\ d \end{array} \right )_L$ & 3 & 2 & 1 & $\frac{1}{3}$ & + \\
$Q_R = \left (\begin{array}{c} u\\ d \end{array} \right )_R$ & 3 & 1 & 2 & $\frac{1}{3}$ & + \\ 
$l_L = \left (\begin{array}{c} \nu \\ e \end{array} \right )_L$ & 1 & 2 & 1 & -1 & + \\ 
$l_R = \left (\begin{array}{c} \nu \\ e \end{array} \right )_R$ & 1 & 1 & 2 & -1 & + \\ 
$U_L$, $U_R$ & 3 & 1 & 1 & $\frac{4}{3}$ & + \\ 
%$U_R$ & 3 & 1 & 1 & $\frac{4}{3}$ & +\\
$D_L$, $D_R$ & 3 & 1 & 1 & $-\frac{2}{3}$ & + \\ 
%$D_R$ & 3 & 1 & 1 & $-\frac{2}{3}$ & +\\ 
$E_L$, $E_R$ & 1 & 1 & 1 & -2 & - \\
%$E_R$ & 1 & 1 & 1 & -2 & -\\ 
$N_L$, $N_R$ & 1 & 1 & 1 & 0 & - \\ 
%$N_R$ & 1 & 1 & 1 & 0 & -\\ 
$H_{RQ}= \left (\begin{array}{c} H_{RQ}^+ \\H_{RQ}^0 \end{array} \right )$ & 1 & 1 & 2 & 1 & + \\
$H_{LQ}= \left (\begin{array}{c} H_{LQ}^+ \\H_{LQ}^0 \end{array} \right )$ & 1 & 2 & 1 & 1 & + \\ 
$H_{Rl}= \left (\begin{array}{c} H_{Rl}^+ \\H_{Rl}^0 \end{array} \right )$ & 1 & 1 & 2 & 1 & - \\ 
$H_{Ll}= \left (\begin{array}{c} H_{Ll}^+ \\H_{Ll}^0 \end{array} \right )$ & 1 & 2 & 1 & 1 & - \\
\end{tabular}}
\caption{{Particle spectrum for lepton-specific LR Universal Seesaw Model.}}
\label{tab:qnumbers}
\end{center}
\end{table}
% 
%\end{widetext}

The scalar sector in this model does not contain any bidoublet fields and hence heavy singlet 
quarks and leptons are necessary for the generation of the quark and lepton masses through 
seesaw-like mechanism. We introduce heavy up and down type quarks given as 
$U_L(3,1,1,\frac{4}{3})$, $U_R(3,1,1,\frac{4}{3})$, and $D_L(3,1,1,-\frac{2}{3})$, $D_R(3,1,1,-\frac{2}{3})$ respectively. Heavy charged 
leptonic states are $E_L(1,1,1,-2)$ and $E_R(1,1,1,-2)$ while heavy neutrino states are given as 
$N_L(1,1,1,0)$ and $N_R(1,1,1,0)$. It is worth noting that while all the $(B-L)$ charged heavy states 
can only have Dirac-like terms, the heavy neutrinos can admit both Dirac and Majorana-like terms. 
Of course one can still argue that the LRS models with triplet Higgs are also lepton 
specific as they do not couple to the quarks. However one must note that the scalar sector 
would then define a significantly different phenomenology from that of the standard triplet 
scenarios, in particular with the absence of a double charged scalar in the spectrum.

The minimal Higgs sector consists of the following fields:
\begin{eqnarray}
H_{RQ}(1,1,2,1)&=&\left (\begin{array}{c}
H_{RQ}^+ \\H_{RQ}^0 \end{array} \right ),~~
H_{LQ}(1,2,1,1)=\left (\begin{array}{c}
H_{LQ}^+ \\H_{LQ}^0 \end{array} \right ),~~ \notag \\
H_{Rl}(1,1,2,1)&=&\left (\begin{array}{c}
H_{Rl}^+ \\H_{Rl}^0 \end{array} \right ),~~
H_{Ll}(1,2,1,1)=\left (\begin{array}{c}
H_{Ll}^+ \\H_{Ll}^0 \end{array} \right ).~~~~~
\end{eqnarray}
where $H_{LQ}$ and $H_{RQ}$ interact specifically with quarks while $H_{Ll}$ and $H_{Rl}$ 
only have leptonic interactions. The $H_{RQ}^0$ and $H_{Rl}^0$ get non-zero VEVs and are 
responsible for breaking the right-handed symmetry. The heavy $W_R$ and $Z_R$ gauge boson 
masses are generated at this scale. The VEVs of the $H_{LQ}^0$ and $H_{Ll}^0$ fields on the other 
hand are the ones responsible for the electroweak symmetry breaking and generation of the $W$ and 
$Z$ boson masses. Since $H_{Rl}^0$ and $H_{Ll}^0$, which are both odd under the $Z_2$ symmetry, 
get non-zero VEVs this could lead to formation of domain walls and destabilize the model. 
This problem is addressed by introducing soft $Z_2$-breaking terms in the scalar potential which 
we discuss later.

The VEV of the Higgs fields are naturally given as (for the universal seesaw
mechanism to work):
\begin{equation}
\left<H_{RQ}^0\right>=v_{RQ},~~\left<H_{Rl}^0\right>=v_{Rl},~~\left<H_{LQ}^0\right>=v_{LQ},~~\left<H_{Ll}^0\right>=v_{Ll},
\end{equation}
with the condition that $v_{LQ}^2+v_{Ll}^2 = v_{EW}^2$. The hierarchy in the VEVs responsible for 
symmetry breaking are arranged as
\begin{equation}
v_{RQ},v_{Rl}>>v_{LQ}>v_{Ll}.
\end{equation}
This ensures a naturally heavy mass for the right-handed gauge bosons which have so far eluded 
any signal at the LHC.

We introduce a lepton-specific $Z_2$ symmetry under which the $E_L, E_R, N_L, N_R, H_{Ll}$ and 
$H_{Rl}$ fields are odd while all other fields are even. This prevents the $Z_2$-odd Higgs fields 
from interacting with the quarks. Table~\ref{tab:qnumbers} has a list of all the particles along with their 
respective quantum numbers.

The covariant derivatives appearing in the kinetic terms of the Lagrangian that lead to interaction 
vertices of the fermions and scalars with the gauge bosons (for all the doublet fields) in this model 
are defined as
\begin{eqnarray}
D_{\mu}Q_L&=& [\partial_{\mu}-i\frac{g_{L}}{2} \tau.W_{L{\mu}}-i\frac{g_V}{6}V_{\mu}]Q_L \nonumber \\
D_{\mu}Q_R&=& [\partial_{\mu}-i\frac{g_{R}}{2} \tau.W_{R{\mu}}-i\frac{g_V}{6}V_{\mu}]Q_R \nonumber \\
D_{\mu}l_L&=& [\partial_{\mu}-i\frac{g_{L}}{2} \tau.W_{L{\mu}}+i\frac{g_V}{2}V_{\mu}]l_L \nonumber \\
D_{\mu}l_R&=& [\partial_{\mu}-i\frac{g_{R}}{2} \tau.W_{R{\mu}}+i\frac{g_V}{2}V_{\mu}]l_R \nonumber \\
D_{\mu}H_R &=& \left[ \partial_{\mu}-i\frac{g_{R}}{2} \tau.W_{R{\mu}}-i \frac{g_V}{2} V_{\mu}\right] H_R \nonumber \\
D_{\mu}H_L &=& \left[ \partial_{\mu}-i\frac{g_{L}}{2} \tau.W_{L{\mu}}-i \frac{g_V}{2} V_{\mu}\right] H_L,
\end{eqnarray}
where $V_{\mu}$ is the $U(1)_{B-L}$ gauge boson and $g_V$ its gauge coupling, while $W_L$, $W_R$ and $g_L$, $g_R$ are the gauge bosons and gauge couplings corresponding to the 
$SU(2)_L$ and $SU(2)_R$ gauge groups respectively. The gauge boson masses can be calculated 
from the kinetic terms for the Higgs boson fields involving the above covariant derivatives. The 
charged gauge boson mass-squared matrix in the basis $(W_R^\pm,W^\pm_L)$ is given as:
\begin{equation}
\begin{bmatrix}
\frac{1}{2}g_R^2(v_{RQ}^2+v_{Rl}^2)&0\\
0&\frac{1}{2}g_L^2(v_{LQ}^2+v_{Ll}^2)
\end{bmatrix}.
\end{equation}
We can clearly see that unlike the case of LRS with bidoublet scalar fields, there is no mixing between 
the two $W$ boson states in this case. The mass of the heavy $W_R$ gauge boson and the SM $W_L$ 
gauge boson states are thus trivially given as:
\begin{equation}
M^2_{W_R^\pm} = \frac{1}{2} g_R^2 (v_{RQ}^2 +v_{Rl}^2),~~~~ 
M^2_{W^\pm} = \frac{1}{2} g_L^2 (v_{LQ}^2 +v_{Ll}^2).
\label{eq:WR}
\end{equation}
The neutral gauge boson mass-squared matrix in the basis $(W_{3R},W_{3L},V)$ is given as:
\begin{equation}
\begin{bmatrix}
\frac{1}{4}g_R^2(v_{RQ}^2+v_{Rl}^2)&0&-\frac{1}{4}g_R g_V(v_{RQ}^2+v_{Rl}^2)\\
0&\frac{1}{4}g_L^2(v_{LQ}^2+v_{Ll}^2)&-\frac{1}{4}g_L g_V(v_{LQ}^2+v_{Ll}^2)\\
-\frac{1}{4}g_R g_V(v_{RQ}^2+v_{Rl}^2)&-\frac{1}{4}g_L g_V(v_{LQ}^2+v_{Ll}^2)&\frac{1}{4}g_V^2(v_{RQ}^2+v_{Rl}^2+v_{LQ}^2+v_{Ll}^2)
\end{bmatrix}.
\end{equation}
This matrix has a zero eigenvalue corresponding to the massless photon state and two other 
non-zero eigenvalues corresponding to the $Z$ and the $Z_R$ bosons. In the limit 
$v_{EW}<<v_{RQ},v_{Rl}$ and keeping only terms upto $v_{EW}^2/v_{RQ(Rl)}^2$, the masses 
of the two massive neutral gauge bosons are given by:
\begin{equation}
M^2_{Z_R} \simeq \frac{1}{2} \left[(g_R^2+g_V^2) (v_{RQ}^2 +v_{Rl}^2)+ \frac{g_V^4(v_{LQ}^2+v_{Ll}^2)}{g_R^2+g_V^2} \right],~~~~M^2_{Z} \simeq \frac{1}{2} (g_L^2 + g_Y^2)(v_{LQ}^2 +v_{Ll}^2),
\label{eq:ZR}
\end{equation}
with the effective SM $U(1)_Y$ gauge coupling given as
\begin{equation}
g_Y=\frac{g_L g_V}{\sqrt{g_L^2+g_V^2}}.
\end{equation}
Quite clearly, in this model the $Z_R$ is heavier than the $W_R$ and therefore a strong limit on the 
$W_R$ mass from experiments would mean an indirect bound exists on the $Z_R$ gauge boson too.

\subsection{Fermion masses and mixings}
\label{sec:fermions}

We now look at the mass of the matter fields in the model. The gauge invariant Yukawa Lagrangian 
respecting the additionally imposed $Z_2$ symmetry in this model is given as:
\begin{eqnarray}
\mathcal{L}_Y&=& \left( Y_{uL}\overline{Q}_{L} \wt{H}_{LQ} U_R + Y_{uR}\overline{Q}_{R} \wt{H}_{RQ} U_L + Y_{dL}\overline{Q}_{L} H_{LQ} D_R + Y_{dR}\overline{Q}_{R} H_{RQ} D_L \right. \notag \\
&+& Y_{\nu L}\overline{l}_{L} \wt{H}_{Ll} N_R + Y_{\nu R}\overline{l}_{R} \wt{H}_{Rl} N_L + Y_{eL}\overline{l}_{L} H_{Ll} E_R + Y_{eR}\overline{l}_{R} H_{Rl} E_L + M_U \ov U_L U_R + M_D \ov D_L D_R \notag \\
&+& \left. M_E \ov E_L E_R + M_{ N} \ov N_L N_R + H.C. \right) + M_{ L} N_L N_L + M_{ R} N_R N_R~~~~
\label{eq:Yukawa}
\end{eqnarray}
where $Y_{iA}$'s are the Yukawa coupling matrices and $M_X$'s are the singlet mass terms 
allowed by gauge symmetry. The conjugated scalar fields are defined as
\begin{equation}
\widetilde H_{L/R} = i \tau_2 H^\ast _{L/R}.
\end{equation}
It is easy to see that the quark and charged lepton mass matrices will consist of off-diagonal terms 
proportional to left and right-handed VEVs while diagonal terms exist for only the heavy fields. Thus 
the quark and charged lepton mass matrices would be very similar in form to the Type-I seesaw 
neutrino mass matrix and all fermions have the same mechanism of mass generation in this framework. 
\subsubsection{Quarks}
The quark masses in this model are obtained by diagonalizing a $6\times6$ mass matrix quite 
similar to what happens in seesaw mechanism. The up quark mass terms in this model can be 
written as
\begin{equation}
\mathcal{L}_u = \begin{pmatrix} \overline{u} & \overline{U}\end{pmatrix}\left( M_u P_L +M_u^T P_R \right) \begin{pmatrix}u\\U\end{pmatrix},
\end{equation}
where
\begin{equation}
M_u = \begin{pmatrix}
0&Y_{uR} v_{RQ} \\ Y_{uL}^T v_{LQ} & M_U
\end{pmatrix}
\end{equation}
is the $6\times6$ up quark mass matrix while $Y_{uL}$, $Y_{uR}$ and $M_U$ are all $3\times3$ 
matrices. The first $3\times3$ block corresponding to the light up-type quark is zero due to the 
absence of a bidoublet field in the scalar spectrum. The off-diagonal terms are obtained from the 
$Y_{uL}$ and $Y_{uR}$ terms of Eqn.~\ref{eq:Yukawa} which involve the mixing of the light and 
heavy states through the Higgs doublet field, while the $M_U$ matrix is the mass term for the heavy 
up-type quarks. For simplicity we will choose all the Yukawa and heavy mass matrices to be diagonal 
in the up sector. This would mean that the CKM mixings will be generated entirely from the down 
sector which is exactly what we do for SM.
 
Similarly the down-type quark mass matrix can be written as
\begin{equation}
M_d = \begin{pmatrix}
0&Y_{dR} v_{RQ} \\ Y_{dL}^T v_{LQ} & M_D
\end{pmatrix},
\end{equation}
where the first $3\times3$ block is again zero due to the absence of a bidoublet scalar while 
the off-diagonal 
blocks arise from the Yukawa couplings. The $M_D$ term is the mass term for the heavy 
down-type quarks. In the down sector too, we keep the right-handed $3\times3$ Yukawa 
matrix $Y_{dR}$ and the $M_D$ matrix to be diagonal while only the left-handed 
Yukawa matrix $Y_{dL}$ is non-diagonal and sufficient to generate the correct CKM mixings for the 
SM quarks.

To diagonalize these non-symmetric matrices we require bi-unitary transformations. For the up-type 
quark mass matrix we have 
\begin{equation}
M_u^{diag} = U_{uL} M_u U_{uR}^{\dagger},
\label{eq:upmass}
\end{equation}
where $U_{uL}$ and $U_{uR}$ are the left and right-handed rotation matrices respectively. Similarly for the down sector
\begin{equation}
M_d^{diag} = U_{dL} M_d U_{dR}^{\dagger}.
\label{eq:downmass}
\end{equation} 
We will get two CKM mixing matrices in this case -- one for the left-handed quarks and another 
for the right-handed quarks, given by
\begin{equation}
U_{L}^{CKM} = U_{uL} U_{dL}^\dagger
\end{equation}
and
\begin{equation}
U_{R}^{CKM} = U_{uR} U_{dR}^\dagger
\end{equation}
respectively. These will be $6\times6$ matrices whose top-left (bottom-right) $3\times3$ block will correspond to the light CKM mixings for ascending (descending) arrangement of eigenvalues by mass. For our choice of parameters and with only the left-handed Yukawa being non-diagonal, the right-handed CKM matrix would be almost diagonal with the mixings being quite small while the left-handed CKM mixings must be the same as the experimentally measured values.

The mixing between the heavy singlet quarks and the SM quarks are determined by the magnitude of the Yukawa terms in comparison to the singlet mass terms. For light quarks and even for the b quark, the Yukawa terms are much smaller than the bare mass term and hence the mixing is very small. For the top quarks though, because of its heavy mass compared to the other SM quarks, the mixings can be quite significant. 

\subsubsection{Charged Lepton}
The charged lepton mass matrix is given as
\begin{equation}
M_e = \begin{pmatrix}
0&Y_{eR} v_{Rl} \\ Y_{eL}^T v_{Ll} & M_E
\end{pmatrix}.
\end{equation}
This is very similar to the quark mass matrix with $Y_{eL}$ and $Y_{eR}$ being the $3\times3$ Yukawa matrices while $M_E$ is the heavy lepton mass matrix. Here we will choose all the matrices to be diagonal to prevent charged lepton flavor violation at the tree level. The mixing between the heavy singlet leptons and the SM charged leptons are almost negligible due to the hierarchical structure of the diagonal and the off-diagonal elements required for generation of correct lepton masses. 

\subsubsection{Neutrino}

The neutrino matrix, on the other hand, would be quite different due to the Majorana-like $M_L$ and $M_R$ terms that could be written for the heavy neutrino states. The neutrino mass matrix in the basis $(\nu_L^\ast,N_R,\nu_R,N_L^\ast)$ is given as
\begin{equation}
\begin{pmatrix} 
0&Y_{\nu L}v_{Ll}&0&0 \\Y_{\nu L}^T v_{Ll}&M_R&0&M_N^T \\ 0&0&0&Y_{\nu R}^T v_{Rl}\\0&M_N&Y_{\nu R} v_{Rl}&M_L
\end{pmatrix}.
\end{equation}
Thus we see that all the fermion masses in this model arise from seesaw-like mass generation mechanism. Hence this model is also popularly known as the universal seesaw model. It is worth noting here that the neutrino mass matrix is actually symmetric (if the Yukawa couplings and heavy mass matrices are symmetric) and can be diagonalized by a simple unitary transformation.

The neutrino mass matrix allows for a number of very unique scenarios in the neutrino sector. Firstly, there is the possibility that only the three left-handed doublet neutrinos are light and everything else is heavy. We explore such a scenario where we get three light neutrinos, three with mass at around electroweak scale and the rest at the TeV scale. We will refer to this scenario as Majorana case for obvious reasons. Similar to the previous cases we will again choose most of the Yukawa and mass matrices here to be diagonal except $Y_{\nu L}$ which is chosen to be a non-diagonal symmetric matrix to explain the experimentally observed Pontecorvo-Maka-Nakagawa-Sakata (PMNS) neutrino mixing matrix elements. The mixing between the light and the heavy states are again quite small here leading to no significant limits from experimental observations. The heavy singlet states though can mix among themselves due to the presence of the $M_N$ term, but due to their masses being at the TeV scale, no observable effects have been discovered so far.

The second case that we can get is when the singlet neutrino Dirac mass term $M_N$ is zero. In this case the neutrino mass matrix becomes block-diagonal with $(\nu_L,N_R)$ and $(\nu_R, N_L)$ bases being the diagonal blocks. In this case we get pseudo-Dirac like states with both the left and right doublet neutrinos being degenerate and light while the heavy singlet states may or may not be degenerate depending upon the choice of parameters. We refer to this scenario as pseudo-Dirac case. If we choose both $M_L$ and $M_R$ to be diagonal and equal, we are forced to choose both $Y_{\nu L}$ and $Y_{\nu R}$ to be non-diagonal. Furthermore, in order to get equal masses for the now light left-handed and right-handed doublet lepton neutral components (neutrinos) we get the condition 
\begin{equation}
Y_{{\nu R}_{ij}} = \frac{v_{Ll}}{v_{Rl}} Y_{{\nu L}_{ij}} \, ,
\label{eq:dn}
\end{equation}
given we take $M_L = M_R$. The mixing between the light and heavy states in each block diagonal sub-matrix are still very small due to the fact that the Yukawa terms are now extremely small compared to the mass terms required to generate the light neutrino masses. The heavy states in this case do not mix as the $M_N$ term is also absent.
 
\subsection{Scalar masses and mixing}
\label{sec:scalars}

The full gauge invariant scalar potential for our model is given as:
\begin{eqnarray}
V(H)&=& \sum_{i=1}^4 \mu_{ii} H_i^\dagger H_i + \sum_{\substack{{i,j=1} \\ {i \leq j}}}^4 \l_{ij} H_i^\dagger H_i H_j^\dagger H_j + \left[\a_1 H_{LQ}^\dagger H_{Ll} H_{RQ}^\dagger H_{Rl}+\a_2 H_{LQ}^\dagger H_{Ll} H_{Rl}^\dagger H_{RQ} \right. \notag \\
&+& \left. \mu_{12}^2 H_{LQ}^\dagger H_{Ll} + \mu_{34}^2 H_{RQ}^\dagger H_{Rl}+ H.C. \right]~~~~
\end{eqnarray}
where
\begin{equation}
H_1 = H_{LQ},~~ H_2 = H_{Ll},~~H_3 = H_{RQ},~~H_4 = H_{Rl}.
\end{equation}
The last two terms involving $\mu_{12}$ and $\mu_{34}$ are responsible for breaking the 
discrete $Z_2$ symmetry softly without introducing any domain walls which could otherwise 
destabilize the model. We minimize this potential and get the following minimization conditions 
\begin{eqnarray}
\mu_{11} &=& 
 \frac{1}{v_{LQ}} \left(\a_{12}^+ v_{Ll} v_{Rl} v_{RQ} % +\a_2 v_{Ll} v_{Rl} v_{RQ} 
 + 2 \l_{11} v_{LQ}^3 + \l_{12} v_{Ll}^2 v_{LQ} + \l_{13} v_{LQ} v_{RQ}^2 + \l_{14} v_{LQ} v_{Rl}^2 - \mu^2_{12} v_{Ll} \right) \, , \notag \\
\mu_{22} &=& 
 \frac{1}{v_{Ll}} \left(\a_{12}^+ v_{LQ} v_{Rl} v_{RQ} %+ \a_2 v_{LQ} v_{Rl} v_{RQ} 
 + \l_{12} v_{Ll} v_{LQ}^2 + 2 \l_{22} v_{Ll}^3 + \l_{23} v_{Ll} v_{RQ}^2 + 
 \l_{24} v_{Ll} v_{Rl}^2 - \mu^2_{12} v_{LQ}\right) \, , \notag \\
\mu_{33} &=& 
 \frac{1}{v_{RQ}} \left(\a_{12}^+ v_{Ll} v_{LQ} v_{Rl} %+\a_2 v_{Ll} v_{LQ} v_{Rl} 
 + \l_{13} v_{LQ}^2 v_{RQ} + \l_{23} v_{Ll}^2 v_{RQ} + 2 \l_{33} v_{RQ}^3 + \l_{34} v_{Rl}^2 v_{RQ} - \mu_{34}^2 v_{Rl} \right) \, , \notag \\
\mu_{44} &=& 
 \frac{1}{v_{Rl}} \left(\a_{12}^+ v_{Ll} v_{LQ} v_{RQ} %+ \a_2 v_{Ll} v_{LQ} v_{RQ} 
 + \l_{14}  v_{LQ}^2 v_{Rl} +\l_{24} v_{Ll}^2 v_{Rl} +\l_{34} v_{Rl} v_{RQ}^2 + 2 \l_{44} v_{Rl}^3 - \mu_{34}^2 v_{RQ} \right) \, , 
\end{eqnarray}
where $ \a_{12}^+ = (\a_1 +\a_2)$.

The Higgs boson spectrum in this case is significantly large and consists of four CP-even states, 
two CP-odd states and two charged Higgs bosons. Two charged goldstone bosons are eaten 
up by the $W_L$ and $W_R$ gauge boson to give them mass while two neutral goldstone states 
give mass to the $Z$ and $Z_R$. The charged Higgs mass-squared matrix in this case is a 
$4\times4$ block diagonal matrix with two blocks of $2\times2$. In the basis 
$(H^+_{LQ},H^+_{Ll},H^+_{RQ},H^+_{Rl})$ the charged Higgs mass-squared matrix is given as:
\begin{equation}
\begin{pmatrix}
\frac{v_{Ll}}{v_{LQ}} \left\{\mu_{12}^2 - \a_{12}^+ v_{Rl} v_{RQ}\right\} & 
 \a_{12}^+ v_{Rl} v_{RQ} - \mu_{12}^2 & 0 & 0 \\
 \a_{12}^+ v_{Rl} v_{RQ} - \mu_{12}^2 & \frac{v_{LQ}}{v_{Ll}} \left\{\mu_{12}^2 - \a_{12}^+v_{Rl} v_{RQ}\right\} & 0 & 0\\
 0 & 0 & \frac{v_{Rl}}{v_{RQ}} \left\{\mu_{34}^2 - \a_{12}^+ v_{Ll} v_{LQ}\right\} & \a_{12}^+ v_{Ll} v_{LQ} - \mu_{34}^2 \\
 0 & 0 & \a_{12}^+ v_{Ll} v_{LQ} - \mu_{34}^2 & \frac{v_{RQ}}{v_{Rl}} \left\{ \mu_{34}^2 - \a_{12}^+ v_{Ll} v_{LQ}\right\}
\end{pmatrix} \, .
\end{equation}
%where $ \a_{12}^+ = (\a_1 +\a_2)$. 
Diagonalizing this matrix we get two goldstone states which are given as
\begin{equation}
G_1^+=\frac{1}{v^2_{Ll}+v^2_{LQ}}\left(v_{LQ},v_{Ll},0,0 \right)^T,~~G_2^+=\frac{1}{v^2_{Rl}+v^2_{RQ}}\left(0,0,v_{RQ},v_{Rl} \right)^T.
\end{equation}
The two physical charged Higgs boson masses are 
\begin{eqnarray}
m^2_{H_1^+} & = & \frac{v^2_{Rl}+v^2_{RQ}}{v_{Rl} v_{RQ}} \left(\mu_{34}^2 - \a_{12}^+ v_{Ll} v_{LQ}\right), ~~m^2_{H_2^+} =\frac{v^2_{Ll}+v^2_{LQ}}{v_{Ll} v_{LQ}} \left(\mu_{12}^2 - \a_{12}^+ v_{Rl} v_{RQ}\right) ,
\label{eq:chhig}
\end{eqnarray} 
with the eigenstates being
\begin{equation}
H_1^+=\frac{1}{v^2_{Rl}+v^2_{RQ}}\left(0,0,-v_{Rl},v_{RQ} \right)^T,~~H_2^+=\frac{1}{v^2_{Ll}+v^2_{LQ}}\left(-v_{Ll},v_{LQ},0,0 \right)^T.
\label{eq:chhig1}
\end{equation}
It is easy to see here that if we choose $\mu_{12}^2=\mu_{34}^2\sim v^2_{EW}$ then the right-handed charged Higgs boson is indeed the lightest state. This is due to the fact that the left-handed charged state has an additional enhancement of $v_{LQ}/v_{Ll}$ except for a very fine-tuned region around
\begin{equation}
\a_1+\a_2 \approx \frac{\mu_{12}^2}{v_{Rl}v_{RQ}}.
\end{equation} 

The CP-odd Higgs boson mass-squared matrix in the basis $({\text{Im}}H^0_{LQ},{\text{Im}}H^0_{Ll},{\text{Im}}H^0_{RQ},{\text{Im}}H^0_{Rl})$ is
\begin{equation}
\begin{pmatrix}
\frac{v_{Ll}}{v_{LQ}} \left(\mu_{12}^2 - \a_{12}^+ v_{Rl} v_{RQ}\right) & 
 \a_{12}^+ v_{Rl} v_{RQ} - \mu_{12}^2 & v_{Ll} v_{Rl} (\a_2-\a_1) & v_{Ll} v_{RQ} (\a_1-\a_2) \\
 \a_{12}^+ v_{Rl} v_{RQ} - \mu_{12}^2 & \frac{v_{LQ}}{v_{Ll}} \left(\mu_{12}^2 - \a_{12}^+ v_{Rl} v_{RQ}\right) & v_{LQ} v_{Rl} (\a_1-\a_2) & v_{LQ} v_{RQ} (\a_2-\a_1)\\
 v_{Ll} v_{Rl} (\a_2-\a_1) & v_{LQ} v_{Rl} (\a_1-\a_2) & \frac{v_{Rl}}{v_{RQ}} \left(\mu_{34}^2 - \a_{12}^+ v_{Ll} v_{LQ}\right) & 
 \a_{12}^+ v_{Ll} v_{LQ} - \mu_{34}^2 \\
 v_{Ll} v_{RQ} (\a_1-\a_2) & v_{LQ} v_{RQ} (\a_2-\a_1) & \a_{12}^+ v_{Ll} v_{LQ} - \mu_{34}^2 & \frac{v_{RQ}}{v_{Rl}} \left(\mu_{34}^2 - \a_{12}^+ v_{Ll} v_{LQ}\right)
\end{pmatrix}.
\end{equation}
This again will have two zero eigenstates corresponding to the two goldstone bosons required for $Z$ and $Z_R$ mass generation. It is also easy to see here that in the case where $\a_1=\a_2$ this CP-odd mass-squared matrix would reduce to the block diagonal charged Higgs boson mass-squared matrix. 

The CP-even scalar Higgs boson mass-squared matrix elements in the basis $({\text{Re}}H^0_{LQ},{\text{Re}}H^0_{Ll},{\text{Re}}H^0_{RQ},{\text{Re}}H^0_{Rl})$ can be expressed in terms of the CP-odd Higgs mass-squared matrix elements as
\begin{equation}
M^2_{{ij,CP-Even}} = M^2_{{ij,CP-Odd}} +2 S_{ij} \l_{ij} v_i v_j,
\end{equation}
where $i,j=1,2,3,4$, $v_{1} = v_{LQ}, v_2=v_{Ll},v_3=v_{RQ},v_{4}=v_{Rl}$ and
\begin{equation}
S_{ij} = \left\{
 \begin{array}{@{}ll@{}}
 2, & {\text{if}}\ i = j \\
 1, & \text{otherwise}.
 \end{array}\right.
\end{equation}
We choose our parameters such that the lightest eigenvalue of this CP-even Higgs 
mass-squared matrix is the one corresponding to the SM-like Higgs with mass of 
125 GeV. This state is consistent with the SM Higgs properties in all its decay channels 
and branching ratios, while all the other states are chosen to be much heavier. Note that
we have implemented the model in \texttt{SARAH} \cite{sarah} and use the 
generated \texttt{SPHENO} \cite{spheno} code to obtain the model spectrum and calculate the 
decay of 
various particles. We have checked that the light Higgs boson of 125 GeV is consistent with
 the expected branching ratios as well as the total decay width of the Standard Model Higgs 
 boson. If we consider 
the $H_1 \rightarrow \gamma \gamma$ decay channel for instance, it gives us a branching ratio of 
$2.27\times 10^{-3}$. We can write the partial decay width as \cite{djouadi}
\begin{eqnarray}
\Gamma_{H_1\rightarrow \gamma \gamma} &=& \frac{G_F \alpha^2 m_{H_1}^3}{128 \sqrt{2} \pi^3} \left| \sum_f N_c Q_f^2 g_{H_1 ff}A_{1/2}^{H_1}\left( \tau_f \right) + g_{H_1 VV} A_1^{H_1}\left( \tau_W \right) \right.\notag \\
&+& \left. \frac{M_W^2 \l_{H_1 H_1^+ H_1^-}} {2 c_W^2 M_{H_1^\pm}^2}A_0^{H_1}\left( \tau_{H_1^\pm} \right) + \frac{M_W^2 \l_{H_1 H_2^+ H_2^-}} {2 c_W^2 M_{H_2^\pm}^2}A_0^{H_1}\left( \tau_{H_1^\pm} \right) \right|^2
\end{eqnarray}
which can be re-expressed as
\begin{equation}
\Gamma_{H_1\rightarrow \gamma \gamma} = \frac{G_F \alpha^2 m_{H_1}^3}{128 \sqrt{2} \pi^3} \left| \sum_f F_{1/2}^{H_1}\left( \tau_f, g_{H_1 ff} \right) + F_1^{H_1}\left( \tau_W,g_{H_1 VV} \right) + \sum_i F_0^{H_1}\left( \tau_{H_i^\pm},\l_{H_1 H_i^+ H_i^-} \right) \right|^2
\label{eq:haa}
\end{equation}
giving us a better idea of the relative contribution from each sector. The benchmark point that we have 
chosen (Table \ref{tab:hcomp}) gives us $\l_{H_1 H_1^+ H_1^-} = 0.034$, 
$\l_{H_1 H_2^+ H_2^-} = 1.20$, 
$M_{H_1^\pm} = 224.7$ GeV, $M_{H_2^\pm} = 6772.4$ GeV. The terms in eq.~{\ref{eq:haa}} are 
thus
\begin{eqnarray}
F_{1/2}^{H_1}\left( \tau_t, g_{H_1 t \overline t} \right)&=&1.835,~~F_1^{H_1}\left( \tau_W,g_{H_1 VV} \right) = -8.324, \notag \\
F_0^{H_1}\left( \tau_{H_1^\pm},\l_{H_1 H_1^+ H_1^-} \right) &=& 0.0013, ~~F_0^{H_1}\left( \tau_{H_2^\pm},\l_{H_1 H_2^+ H_2^-} \right) = 3.6 \times 10^{-5}, \notag
\end{eqnarray} 
where the four contributions are from the top quark, $W$ boson and the two charged Higgs states respectively. Note that the contributions from the charged Higgs bosons are orders of magnitude lower compared to the top quark and gauge boson contributions and do not affect the 
$H_1\rightarrow \gamma \gamma$ branching ratio. 

The lightest charged and pseudo-scalar 
Higgs boson masses come out to be around a few 100 GeV while the heavier ones are around a 
few TeV. In Table~\ref{tab:hcomp} we give a list of the physical Higgs boson masses and the respective 
eigenstates for a sample benchmark point. Note that unlike the case of 2HDM models, here 
only the pseudoscalar and charged Higgs bosons are light while all other CP even scalars turn out to 
be very heavy. In addition both the light pseudoscalar and charged scalar are admixtures of the 
right sector scalar doublets. 
\begin{table}[h!]
\begin{center}
{\renewcommand{\arraystretch}{1.5}
\begin{tabular}{C{1.8cm}|C{1.8cm}|C{12.0cm}} \hline 
{Particle} & {Mass (GeV)} & {Eigenstate} \\
\cline{1-3}
$H_1$ & 125.2 & $0.996~{\text{Re}}( H^0_{LQ})-0.008~{\text{Re}}(H^0_{RQ}) + 0.080 ~{\text{Re}}(H^0_{Ll})+0.019~{\text{Re}}(H^0_{Rl})$ \\
$H_2$ & 3386.1 & $-0.0209 ~{\text{Re}}( H^0_{LQ})-0.381~{\text{Re}}(H^0_{RQ}) + 0.001 ~{\text{Re}}(H^0_{Ll})+0.924 ~{\text{Re}}(H^0_{Rl})$ \\
$H_3$ & 5638.1 & $0.001 ~{\text{Re}}( H^0_{LQ})-0.924~{\text{Re}}(H^0_{RQ}) + 0.008 ~{\text{Re}}(H^0_{Ll})-0.381~{\text{Re}}(H^0_{Rl})$ \\
$H_4$ & 6772.5 & $0.080~{\text{Re}}( H^0_{LQ})-0.007~{\text{Re}}(H^0_{RQ}) - 0.997 ~{\text{Re}}(H^0_{Ll})+0.004~{\text{Re}}(H^0_{Rl})$ \\
$A_1$ & 214.8 & $0.001~{\text{Im}}(H^0_{LQ})-0.707~{\text{Im}}(H^0_{RQ}) - 0.010 ~{\text{Im}}(H^0_{Ll})+0.707 ~{\text{Im}}(H^0_{Rl})$ \\
$A_2$ & 6772.7 & $-0.080 ~{\text{Im}}(H^0_{LQ})-0.007 ~{\text{Im}}(H^0_{RQ}) + 0.997 ~{\text{Im}}(H^0_{Ll})+0.006~{\text{Im}}(H^0_{Rl})$ \\
$H^+_1$ & 224.7 & $-0.707 H^+_{RQ}+0.707 H^+_{Rl}$\\
$H^+_2$ & 6772.4 & $0.080 H^+_{LQ} - 0.997 H^+_{Ll}$\\
\cline{1-3}
\end{tabular}}
\caption{{Scalar Eigenstates for $\a_1 = -0.2,~\a_2=0.1,~\l_{11}=0.168,~\l_{12}=0.8,~\l_{13}=0.05,~\l_{14}=-0.1,~\l_{22}=0.5,~\l_{23}=0.1,~\l_{24}=0.1,~\l_{33}=0.2,~\l_{34}=0.1,~\l_{44}=0.1,\mu_{12}^2=2.5\times10^4,~\mu_{34}^2=2.5\times10^4$, $v_{RQ} = v_{Rl}=6.0$ TeV, $v_{LQ}=173.4$ GeV, $v_{Ll}=14$ GeV.}}
\label{tab:hcomp}
\end{center}
\end{table}
To check whether the Higgs potential is stable for our choice of benchmark points, we considered 
the check on the co-positive conditions for stability of the potential when the couplings are negative \cite{jpt}. The condition for the stability of the potential for the negative coupling $\l_{14}$ is given by
\begin{equation}
\l_{11} \geq 0,~~\l_{44} \geq 0,~~\l_{14}\geq -\sqrt{\l_{11} \l_{44}} \, ,
\end{equation}
which are easily satisfied. We checked the condition for $\alpha_1$ numerically, by constructing 
the principal sub-matrices and found that it satisfies the criteria for stability as well.

\section{Phenomenological Implications}
\label{sec:pheno}

%It is worth pointing out here that to generate the model and study the spectrum we have implemented 
%the model into \texttt{SARAH} \cite{sarah} and use the generated \texttt{SPHENO} \cite{spheno} 
%code to analyze the spectrum and decay probabilities of the various particles in our model.
%%

We now look at the phenomenological implications of this model, {\it viz.} the allowed parameters, 
experimental constraints and unique signals of this model which may be studied at the colliders. 
As has been discussed before, almost all of the matrices are taken to be diagonal except the 
$Y_{dL}$ and $Y_{\nu_L}$ matrices. These are necessarily off-diagonal in order to generate 
the CKM and PMNS mixing. Unlike SM where the Yukawa couplings can range from 
$10^{-6}$ to 1, this model requires a much smaller range of Yukawa couplings ranging from 
$10^{-3}$ to 1 for all the charged particles. In general we have chosen
%\begin{eqnarray}
%Y_{u(L,R)}^{11} &\sim & Y_{d(L,R)}^{11} \sim Y_{e(L,R)}^{11} \approx 10^{-2},~~
%Y_{u(L,R)}^{22} \sim Y_{dR}^{22} \sim Y_{e(L,R)}^{22} \approx 10^{-1},~~
%Y_{u(L,R)}^{33} \sim Y_{dL}^{33} \sim Y_{eR}^{33} \approx 1,~~~
%\end{eqnarray}
\begin{align}
Y_{u(L,R)}^{11} \sim Y_{d(L,R)}^{11} \sim Y_{e(L,R)}^{11} \approx 10^{-2} , \,\, &
Y_{u(L,R)}^{22} \sim Y_{dR}^{22} \sim Y_{e(L,R)}^{22} \approx 10^{-1} , &
Y_{u(L,R)}^{33} \sim Y_{dL}^{33} \sim Y_{eR}^{33} \approx 1 ,
\end{align} 
while the other elements in the $Y_{dL}$ matrix are of the order of $10^{-3}$. We further 
choose $Y_{dR}^{33}=0.023$ and $Y_{eL}^{33} =0.26$ so that the third generation heavy fermion 
masses are all of the order of a few TeV. With this kind of a Yukawa structure we can easily get 
the correct masses of all the fermions by choosing appropriate values of the heavy masses. The 
left-handed CKM matrix elements are obtained entirely in the down sector similar to the SM, 
while the right-handed down quark mixings are very small due to the diagonal structure of the 
$Y_{dR}$ matrix. The VEV's are taken to be
\begin{equation}
v_{RQ}= v_{Rl} = 6.0 {\text{ TeV}},~~ v_{LQ} = 173.4 {\text{ GeV}},~~ v_{Ll} = 14 {\text{ GeV}}.
\end{equation} 
The bare masses of the singlet heavy vector-like fermions are chosen accordingly 
so as to get the correct masses for the SM-like fermions. Here we note that since the third 
generation fermions are the heaviest followed by the second generation and then the first 
generation fermion masses, the reverse order is generally followed by the vector-like singlet 
fermion masses. For each type of fermions (up quark, down quark and charged leptons), our 
choices are such that the third generation vector-like fermion is the lightest while the first generation 
is the heaviest. This can be understood easily as in the seesaw formula the mass of the light state 
is inversely proportional to the heavy mass in the seesaw matrix for the same value of off-diagonal 
terms. Though it is not strictly valid for this case as the off-diagonal Dirac masses are also higher 
for the third generation, we choose our Yukawa couplings so that the third generation vector 
fermions are indeed lightest. In Table~\ref{tab:params} we list the mass of all the new fermions in our model. 
With the strong sector exotic quarks and charged leptons having masses above 3.5 TeV, it would be 
quite impossible to observe any signals for these fermions at the current LHC energies. However 
they could be more copiously produced at future 100 TeV machines such as the FCC-hh 
collider \cite{Mangano:2017tke}. 

\begin{table}[h!]
\begin{center}
{\renewcommand{\arraystretch}{1.5}
\begin{tabular}{C{2.8cm}|C{3.1cm}|C{3.0cm}|C{3.2cm}|C{3.1cm}} \hline
\multirow{2}{*}{Up-type Quark} & \multirow{2}{*}{Down-type Quark} & \multirow{2}{*}{Charged Lepton} & \multicolumn{2}{c}{Neutrino} \\
\cline{4-5}
& & & Majorana & Pseudo-Dirac \\
\cline{1-5}
$M_{T} =4.51$ TeV, $M_{C} =6.17$ TeV, $M_{U} =30.0$ TeV & $M_{B} =3.97$ TeV, $M_{S} =10.4$ TeV, $M_{D} =17.2$ TeV & $M_{E_3} =6.13$ TeV, $M_{E_2} =9.92$ TeV, $M_{E_1} =12.3$ TeV & $M_{\nu_4} =136$ GeV, $M_{\nu_5} =258$ GeV, $M_{\nu_6} =317$ GeV, $M_{\nu_7} =9.07$ TeV, $M_{\nu_8} =9.13$ TeV, $M_{\nu_9} =9.16$ TeV, $M_{\nu_{10}} =11.06$ TeV, $M_{\nu_{11}} =11.1$ TeV, $M_{\nu_{12}} =11.2$ TeV & $M_{\nu_4} =200.0$ GeV, $M_{\nu_5} =300.0$ GeV, $M_{\nu_{6}} =400.0$ GeV \\
\cline{1-5}
\end{tabular}}
\caption{{Fermion masses.}}
\label{tab:params}
\end{center}
\end{table}
The mixing between the heavy singlet-like states and the light SM-like states are 
very low ($\lesssim 1 \% $) except for the top sector which behaves quite differently. 
To get the correct top quark mass we need to take $M_u^{33}$ to be quite small to be 
around 350 GeV. The heavy top partner mass almost entirely comes from the right handed 
top quark contribution and hence the right-handed CKM mixing of the top quark is almost 
entirely coming from the heavy singlet top partner. Thus the decay of the heavy 
gauge bosons which belong to the $SU(2)_R$ do not couple with the same strength to the 
third generation SM quarks as they do to the first two generations. This effect is clearly 
visible in the $W_R$ decay modes where its branching ratio $ W_R^+ \to t \bar{b}$ is significantly 
suppressed ($\sim$ 0.2\%) while for the first two generation light quarks it is around 33\% each. 
\begin{figure}[b!]
\centering
\includegraphics[width=4.8in,height=3.5in]{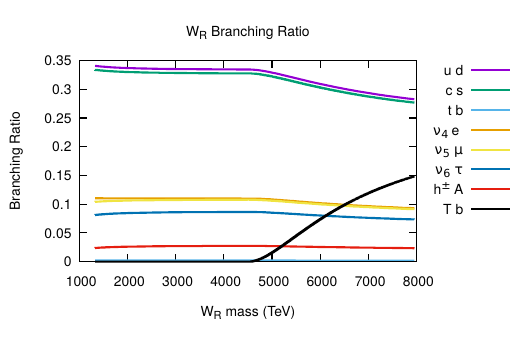}
\caption{{Branching ratio for $W_R$ boson as a function of its mass.}}
\label{fig:wrbr}
\end{figure}
This in turn would make the bounds on the $W_R$ gauge boson much stronger from existing 
dijet data than that of conventional LRS models which have slightly lower branchings into light 
jets. The current bound on a heavy SM-like $W'$ from LHC is 2.6 TeV \cite{dijet}. At a mass of 2.6 TeV, a SM-like $W'$ goes into light quarks with a branching ratio of 47.6\%. The $W_R$ in our case has a 66\% branching ratio into light quarks and therefore the limits would be stronger. Using the experimental bound on cross-section$\times$branching ratio ($\sigma \times BR$) we get 
a lower bound on $W_R$ mass of 2.75 TeV in our model. For our choice of benchmark points, 
the mass of $W_R$ boson comes out to be $\sim 4$ TeV and is safe from dijet bounds. It 
is also allowed from heavy neutrino searches \cite{Sirunyan:2018pom}, provided the heavy neutrino mass is not very heavy. We show the $W_R$ decay modes and the branching probabilities as a 
function of its mass in Fig. \ref{fig:wrbr}.

\begin{figure}[t]\centering
\includegraphics[width=3.2in]{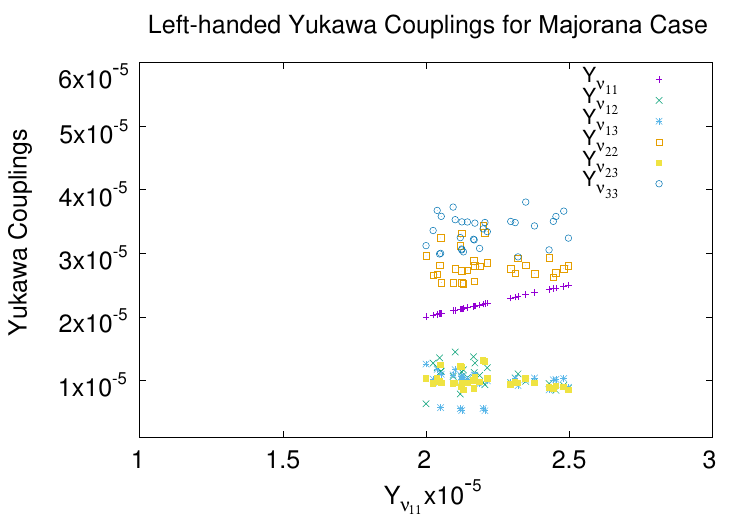} 
\includegraphics[width=3.2in]{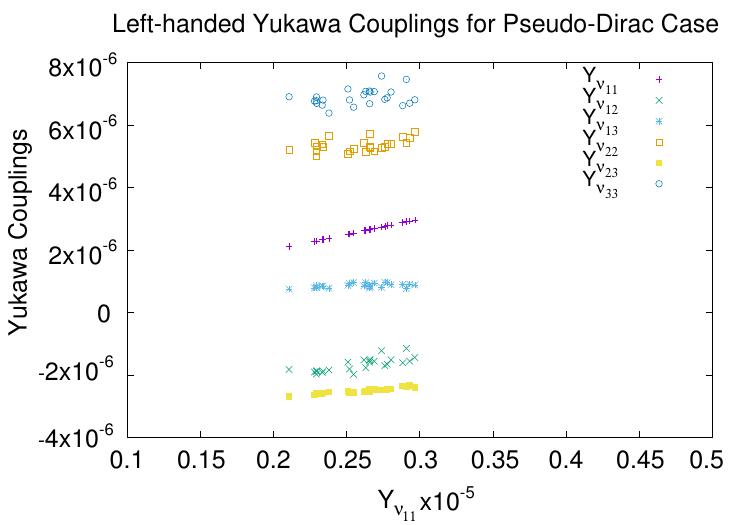}
\caption{{ (a) Benchmark points satisfying the neutrino masses and PMNS mixings for the Majorana 
case, (b) Benchmark points satisfying the neutrino masses and PMNS mixings for the Pseudo-Dirac 
case.}}
\label{fig:Yukawa}
\end{figure}

In the neutrino sector there are two specific scenarios (Majorana and pseudo-Dirac) as has been 
discussed earlier. Here we will only discuss the case of normal hierarchy for the neutrino 
masses.\footnote{It is worth noting that an arrangement for inverted hierarchy of the neutrino masses
is equally possible in our model, which we have not considered here.} 
For the Majorana case, we fit the experimental data for neutrino oscillation parameters \cite{neut-osc} 
with the variations being within the $3\sigma$ range of their respective central values obtained in the
global fits. We have listed the values used for the fit in Table~\ref{tab:numix} while scanning the 
parameter space of our model.
\begin{table}

\begin{center}

{\renewcommand{\arraystretch}{1.5}

\begin{tabular}{||C{9.0cm}|} \hline

 7.03$\times 10^{-5}~\text{eV}^2$ $<\Delta m_{21}^2<$ 8.09$\times 10^{-5}~\text{eV}^2$\\ \hline

 2.407$\times 10^{-3}~\text{eV}^2$ $<\Delta m_{31}^2<$ 2.643$\times 10^{-3}~\text{eV}^2$\\ \hline

 $0.271<\sin^2{\theta_{12}}<0.345$ \\ \hline

 $0.385<\sin^2{\theta_{23}}<0.635$ \\ \hline

 $0.01934<\sin^2{\theta_{13}}<0.02392$ \\ \hline

 $U_{PMNS}$ $$ \begin{pmatrix}

0.800 \rightarrow 0.844 & 0.515 \rightarrow 0.581 & 0.139 \rightarrow 0.155 \\

0.229 \rightarrow 0.516 & 0.438 \rightarrow 0.699 & 0.614 \rightarrow 0.790 \\

0.249 \rightarrow 0.528 & 0.462 \rightarrow 0.715 & 0.595 \rightarrow 0.776\end{pmatrix} $$ \\ \hline

\end{tabular}}

\caption{{Experimental $3\sigma$ ranges for light neutrino parameters.}}

\label{tab:numix}

\end{center}

\end{table}We choose all the matrices to be diagonal except $Y_{\nu L}$ which is a symmetric matrix. We choose 
\begin{eqnarray}
M_R = M_L = {\text{Diag}}\left(10^4,10^4,10^4\right),~~M_N = {\text{Diag}}\left(10^3,10^3,10^3\right),
\end{eqnarray}
while the elements of the Yukawa coupling matrix $Y_{\nu {R_{ii}}} \sim 0.1$ and 
$Y_{\nu_{L_{ij}}} \sim 10^{-5}$. This choice gives us our desired neutrino masses with the three 
light neutrino states lying between 0.001 to 0.05 eV while the next three heavy states have masses 
around a few 100 GeV. The rest of the physical states have mass around 10 TeV. 
The three light neutrino physical states are almost entirely from the three generations of 
$\nu_L$ and the next three (masses of a few 100 GeV) come 
mostly from $\nu_{R}$. This is very similar to Type-I seesaw in conventional LRS models and 
would give very similar phenomenology with same-sign lepton signals and neutrinoless double-beta 
decay. However the modified scalar sector interaction with the heavy neutrinos lead to much 
different collider signals which we shall discuss later.
The much heavier eigenstates which would be beyond the reach of current accelerator energies 
are a mixture of $N_L$ and $N_R$. In Fig.~\ref{fig:Yukawa}(a) we show the allowed parameters which 
gives us the correct neutrino mass-squared differences and the correct PMNS mixing angles for 
normal hierarchy. We can see that $Y_{\nu_{33}}$ is indeed the largest owing to $\nu_3$ being the 
heaviest in normal hierarchy case, while $Y_{\nu_{13}}$ is the smallest in magnitude as required to 
explain the small value of the mixing angle $\theta_{13}$.

The $12\times12$ neutrino mass matrix is symmetric and hence it can be 
diagonalized with a simple unitary transformation. The first $3 \times 3$ block 
corresponds to the three light neutrinos and 
satisfies the experimental 3$\sigma$ bounds of the PMNS matrix. The other mixing of the 
light neutrinos with the heavier ones are extremely small with $\sin \theta_{ij} \lesssim 10^{-8}$ 
where $\theta_{ij}$ is the mixing angle between the heavy and light states. Hence there are 
no bounds coming from lepton flavor changing processes. The three mass eigenstates with 
masses of a few 100 GeV are almost the same as the flavor eigenstates of $\nu_{R_i}$ with 
small mixing ($\sim 1\%$) with $N_{L_i}$. The six heavier states of masses around 10 TeV 
almost equally constitute of $N_{L_i}$ and $N_{R_i}$, where $i=1,2,3$. Table~\ref{tab:params} gives 
a list of all the neutrino masses in this scenario for a particular benchmark point. 

In the pseudo-Dirac neutrino case the singlet neutrino mixing term $M_N$ is taken to be zero. 
Then we choose our parameters as $M_R = M_L = {\text{Diag}}\left(200,300,400\right)$. To get 
the correct light neutrino masses and mixings for this choice of $M_L$ and $M_R$, we are forced 
to choose both $Y_{\nu L}$ and $Y_{\nu R}$ to have non-zero off-diagonal elements (still being 
symmetric matrices). Now we get two block diagonal $6 \times 6$ 
symmetric matrices each of whose three light eigenvalues should be equal and satisfy the 
experimentally observed mass-squared differences for them to form pseudo-Dirac-like states. 
Using Eqn.~\ref{eq:dn} along with the observed mass-squared differences and mixing constraints 
 gives us the following choice of the matrix elements: $Y_{\nu {L_{ij}}} \sim 10^{-6}$ and $Y_{\nu {R_{ij}}} \sim 10^{-9}$. 
The neutrino mixings in this case again have to satisfy the experimental PMNS mixing limits and 
have to be the same for both the sectors. This is easily satisfied by using the condition given 
in eq.~\ref{eq:dn}. Fig.~\ref{fig:Yukawa}(b) gives some allowed parameters for this case which 
satisfy the neutrino mass-squared differences and the mixing angles for normal hierarchy. As 
observed in the previous case, we find that $Y_{\nu_{33}}$ is usually the largest while the 
elements for $Y_{\nu_{13}}$ are the smallest for most of the points. 

The mixing between the states of $\nu_L$ and $\nu_R$ which are now of equal masses in this 
scenario are not quite as small as the previous case. The mixing angle $\theta$ between 
two light states of equal masses are typically such that $\sin\theta \sim 10^{-2}$. The heavy 
states again have negligibly small mixing with the light states like in the previous case. Note that 
as we have taken $M_N =0$ there is no mixing between the heavy states and they are purely 
compositions of $N_{L_i}$ or $N_{R_i}$. As a result their decay width in this case becomes very 
small and with suitable choice of parameters it may lead to observable displaced vertex signals. 
It is worth noting that when the charged Higgs is lighter than the heavy neutrino states, it becomes 
the primary channel of decay and therefore can provide for interesting signal channels for 
the model at collider experiments, which we discuss later in more detail.

%We have listed the sub-TeV heavy 
%Majorana neutrino decay widths and the most important decay channels in Tab.~\ref{tab:nudks} which
%clearly shows the very small decay widths of many of the heavy neutrinos which mostly decay into 
%three body final states which are charged and may be observed in the experiments.

\subsection{Experimental Constraints}
\label{sec:expconst}

\begin{figure}[h!]\centering
\includegraphics[width=3.2in]{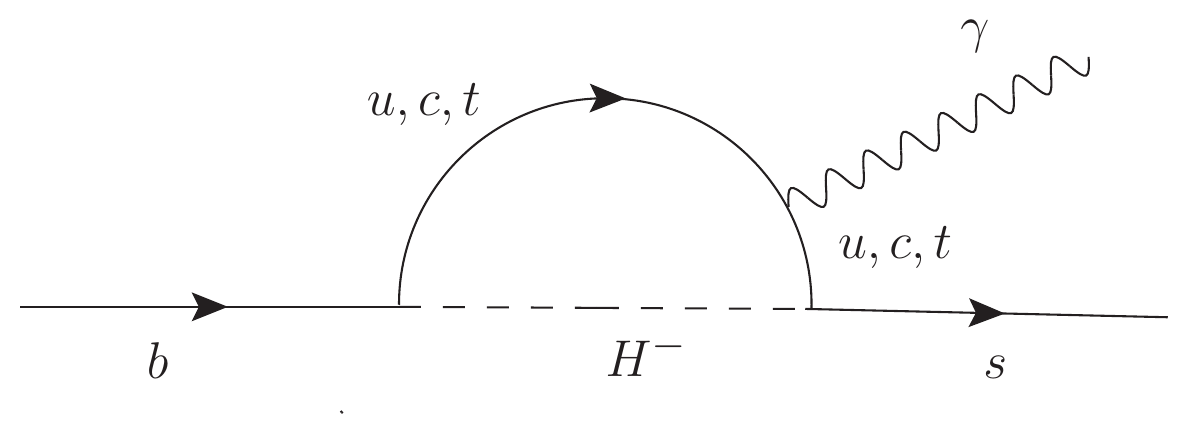}
\caption{{ $b\rightarrow s\gamma$ through charged Higgs.}}
\label{fig:bsg}
\end{figure}

The scalar sector of the model discussed in this paper may be considered as a left-right extension of the lepton specific two Higgs doublet model (2HDM). As such there are a number of flavor constraints which restrict the parameter space of the 2HDM. Most stringent of these constraints come from the $b\rightarrow s \gamma$ process and constrains the charged Higgs mass to $m_{H^\pm}>460$ GeV \cite{Hussain} in Type III 2HDM. The main process responsible for $b\rightarrow s \gamma$ in 2HDM is given in fig~\ref{fig:bsg}. In our case though this process is present, the lighter charged Higgs boson of mass around 200 GeV actually corresponds to the right-handed charged Higgs boson as can be seen from eq.~\ref{eq:chhig1}. As the right-handed down type Yukawa coupling matrix is diagonal in this model, the CKM mixings are really small. This results in a much weaker bound on the lightest charged Higgs boson mass in this case. There are no significant bounds from the flavor observable 
on the pseudoscalar mass in the lepton-specific 2HDM and hence there are no bounds in this 
model as well.

There is a bound on the 2HDM pseudoscalar Higgs boson mass from the single production and associated production of the pseudoscalar decaying into two $\tau$ final state \cite{atlas}. This gives a lower limit on the pseudoscalar mass as a function of the $\sigma \times BR(A\rightarrow\tau \tau)$ for both the single and the associated production mode. For both these production channels the important couplings would be $A_1 q \ov{q}$ where $q$ is a quark. To get significant production, the third generation quarks are the most important but here again the couplings of the pseudoscalar with the third generation quarks will be much weaker than in the case of 2HDM. This is because the coupling here would be 
\begin{equation}
f^L_{Au\ov{u}(Ad\ov{d})}=Y^{33}_{uL(dL)}\times Z_{Ai}^L\times Z_{uR(dR)}^{Qq}
\label{eq:Acoup}
\end{equation}
where $Z_{Ai}^L$ is the amount of $H^0_{LQ}$ contained in the eigenstate of $A$ and $Z_{uR(dR)}^{qQ}$ is the mixing of the right-handed heavy and the light up-type (down-type) quarks. Similar formula can be written for the right-handed pseudoscalar coupling with two quarks with $L\leftrightarrow R$ in Eq.~\ref{eq:Acoup}. The light pseudoscalar in this model is coming from the right-handed doublets and its couplings with the third-generation quarks come out to be much weaker than the 2HDM case. So for our model this limit will not be applicable because the production cross-section of the pseudoscalar will be much smaller than in 2HDM.

\subsection{New Collider Signals}
\label{sec:colliders}

This model can lead to a number of interesting new signals at accelerator experiments. 
We primarily focus in mentioning the ones from the scalar sector in the form of charged 
Higgs as well as the heavy Majorana neutrinos which can be accessible to the current run 
of the Large Hadron Collider (LHC). Note that the other exotics such as the heavy 
quarks and leptons are beyond the reach of LHC because of their extremely heavy masses.

To highlight the signals for the model at LHC we choose two representative points in the 
model parameter space as {\bf BP1} and {\bf BP2} and list them in Table \ref{tab:nudks}. 
Unlike other models for heavy neutrinos including left-right symmetric models we find that
the decay modes of the heavy Majorana neutrinos, as listed in Table \ref{tab:nudks}, are 
quite different. Note that the heavy neutrino decays are again driven by their composition 
and therefore could be either singlet dominated or even $SU(2)_R$ doublet dominated. 
As the Yukawa couplings $Y_{\nu R_{ii}} \sim 0.1$ we find that the dominantly right-sector charged 
Higgs which is light, would couple to the heavy Majorana neutrinos which have dominant right-handed
components as well as singlet components over the left-handed one's. This plays a crucial role 
in deciding the decay of the heavy neutrinos in the model. The heavy neutrinos prefer to decay 
via the off-shell charged Higgs while the subleading contributions come from the decay via 
off-shell $W_R$. Although both the mediating particles would contribute, a quick look at the 
decay probabilities in Table \ref{tab:nudks} shows the absence of the leptonic modes 
for {\bf BP1} which are 
highly suppressed. This indicates that the decay is driven by the off-shell charged Higgs 
over the much heavier $W_R$ gauge boson. The challenge for observing signals for these 
heavy neutrinos would be dictated by the production mechanism. At LHC, it would mean that 
they could be produced via exchange of $W_R$ and $Z_R$ in the $s$-channel. This 
would give a resonant production of the heavy neutrinos and therefore the dominant 
channel. So we can produce the heavy neutrinos as 
$$ \bullet \hspace*{0.5cm} p \, p \rightarrow W_R^{\pm} \rightarrow \ell^{\pm}_i \,\, \nu_j $$ 
$$ \bullet \hspace*{0.5cm} p \, p \rightarrow Z_R \rightarrow \nu_j \,\, \nu_j $$
where $i=1,2,3$ and $j=i+3$.

\begin{table}[!ht]
\begin{center}
{\renewcommand{\arraystretch}{1.5}
\begin{tabular}{C{1.4cm}|C{3.5cm}|C{3.2cm}|L{5.6cm}} \hline
\multicolumn{2}{c|}{Particle} & Width (GeV) & \multicolumn{1}{c}{Important Decay Channels} \\
\cline{1-4}
\multirow{3}{2.0cm}{{\bf BP1}} 
& $M_{\nu_4}=136.4$ GeV &$7.12\times 10^{-9}$ & $\nu_4 \to e^\pm\,jj \sim$ 99.3\% \\
\cline{2-4}
& $M_{\nu_5}=258.3$ GeV & $4.57\times 10^{-4}$ & $\nu_5 \to \mu^\pm\,H_1^\mp \sim$ 100\% \\
\cline{2-4}
& $M_{\nu_6}=317.0$ GeV &$2.97 \times 10^{-3}$ & $\nu_6 \to \tau^\pm\,H_1^\mp \sim$ 100\% \\
\cline{2-4}
& $M_{H_1^\pm}=224.7$ GeV &$4.6 \times 10^{-4}$ & $H_1^\pm \to \nu_4 \,e^\pm \sim$ 99\% \\
\cline{1-4}
\multirow{3}{2.0cm}{{\bf BP2}} 
& $M_{\nu_4}=317.0$ GeV & $2.11\times 10^{-3}$ & $\nu_4 \to \e^\pm\,H_1^\mp \sim$ 100\% \\
\cline{2-4}
& $M_{\nu_5}=550.9$ GeV &$3.03\times 10^{-2}$ & $\nu_5 \to \mu^\pm\,H_1^\mp \sim$ 100\% \\
\cline{2-4}
& $M_{\nu_6}=837.6$ GeV & $1.04 \times 10^{-1}$ & $\nu_6 \to \tau^\pm\,H_1^\mp \sim$ 100\% \\
\cline{2-4}
& $M_{H_1^\pm}=224.7$ GeV &$5.78 \times 10^{-6}$ & $H_1^+ \to t \, \bar{b} \sim$ 92.9\% \\
\cline{1-4}
\end{tabular}}
\end{center}
\caption{Representative benchmark points of the particle spectrum where the massive 
neutrino states are Majorana type. We also illustrate the dominant decay modes 
of the lightest charged Higgs and heavy neutrino states. The heavy gauge bosons for 
both {\bf BP1} and {\bf BP2} are same with $M_{W_R}=4$ TeV and $M_{Z_R}=4.7$ TeV.}
\label{tab:nudks}
\end{table}

We find that a $W_R$ of mass 4 TeV consistent with current experimental limits, has a 
combined branching of nearly $30\%$ to decay to a SM charged lepton and heavy neutrino, 
with the dominant mode of the three being the decay to the 
first 2 generations $\sim11\%$. Similarly a $Z_R$ of mass around 4.8 TeV has a 
combined $\sim 20-22\%$ branching probability to decay in the pair of $\nu_4, \, \nu_5, \, \nu_6$ for {\bf BP1} and {\bf BP2}, respectively. Resonant production of heavy neutrinos 
can be useful to have appreciable rates of production \cite{Huitu:2008gf} without depending on the active-sterile 
mixing parameter in the neutrino sector. As the decay in Table \ref{tab:nudks} suggests, the heavy neutrino decays 
to a single flavor charged lepton in association with jets with a $100\%$ branching probability for {\bf BP1}, 
leading to same-sign dilepton signals with jets in the final state provided the charged Higgs in the model is heavier. In fact one can also have the interesting signal where 
\begin{align*}
 p \, p \to \nu_5 \nu_5 \to \mu^\pm \, \mu^\pm \, H_1^\mp \, H_1^\mp \to 2 \mu^\pm + 2 e^\mp + 4 \, j \,\,\, . 
\end{align*}
The pair production cross section at LHC for the heavy neutrinos with $\sqrt{s}=13$ TeV are 
\begin{align*}
{\bf BP1}: && \sigma(pp \to \nu_4 \nu_4) = 0.121 \, {\rm fb}, && \sigma(pp \to \nu_5 \nu_5) = 0.101 \, {\rm fb}, && \sigma(pp \to H_1^+ H_1^-) = 5.27 \, {\rm fb} \,\,\, .
\end{align*}
The $\nu_4$ production gives the familiar same-sign lepton signal
$$ p \, p \to \nu_4 \nu_4 \to 2 e^\pm + 4 \, j \,\,\, .$$

By slightly changing our parameters we get a spectrum where all the heavy neutrinos are 
actually heavier than the lightest charged Higgs represented by {\bf BP2} as shown in 
Table~\ref{tab:nudks}. We have only modified the neutrino sector making sure that the 
new set of parameters are still consistent with the neutrino oscillation data, while keeping the 
other sectors almost same as before. In this case where all the Majorana neutrinos are 
heavier than the lightest charged Higgs which in turn is heavier than the top quark, a very 
unique and different signal is produced. With no heavy neutrino decay available to the 
charged Higgs, it decays to the quark final states with the dominant channel being 
$H_1^{+} \to t \, \bar{b}$ (see Table \ref{tab:nudks}). Now as the charged Higgs comes 
from the decay of a heavy Majorana neutrino then one gets an interesting signal where 
one has same-sign leptons as well as same-sign top quark in the final state. This 
is completely free from any SM background and would be an unique signal 
for discovery. Thus we have for example 
$$ p \, p \to \nu_4 \nu_4 \to e^\pm \, e^\pm \, H_1^\mp \, H_1^\mp \to 2 e^\pm + 2 \, \bar{t}/t 
+ 2 \, b/\bar{b}$$
when the $\nu_4$ is pair produced. Again one gets $2\mu^\mp \, + 2 t/\bar{t} + 2\bar{b}/b$ when 
$\nu_5$ is pair produced. The total cross section for the pair production of the heavy neutrino pairs at LHC with $\sqrt{s}=13$ TeV are
\begin{align*}
{\bf BP2}: && \sigma(pp \to \nu_4 \nu_4) = 0.096 \, {\rm fb} \, , && \sigma(pp \to \nu_5 \nu_5) = 0.076 \, {\rm fb} \,\,\, .
\end{align*}
Subsequent decay probabilities for heavy neutrino decay as well as charged Higgs decay 
are almost 100\%. To compare it with the SM background, there are no subprocesses 
that can contribute directly to similar final states. Due to charge mismeasurements we can 
consider the process $pp \to e^+ e^- t \bar{t} b\bar{b}$ as a possible background. The cross 
section for this process at LHC with $\sqrt{s}=13$ TeV is around 0.16 fb. The charge 
mismeasurement probabilities are much below $10^{-3}$ and therefore this would hardly 
give any event even with an integrated luminosity of 3000 fb$^{-1}$. The signal would still 
yield a handsome 516 events for {\bf BP2} where we add the 
contributions for both the $\nu_4$ and $\nu_5$ channels in case of {\bf BP2}. 
Although the signal would be difficult to observe in the near future at both ATLAS and CMS, 
but with the very high luminosity option at LHC, it would be a very unique channel to 
observe. 

For the charged Higgs ($H_1^{\pm}$) lighter than the top quark it decays to the 
light quarks thus giving a more conventional signal of same-sign leptons with multiple jets. 
However a marked difference is the absence of SM $W$ and $Z$ boson in the decay 
cascades of the heavy neutrino decay. These modes become available for the scenario 
with pseudo-Dirac heavy neutrinos. Similarly, the single production of the heavy 
neutrino through $W_R$ resonance would lead to a signal with same-sign lepton along 
with a top and bottom quark, where $\sigma(pp \to W_R)=3.2$ fb for $M_{W_R}=4$ TeV. 
We leave a much more detailed signal analysis of the collider signals of the model for 
future work and focus on pointing out the interesting signals that one can expect to 
observe at LHC here.

 In addition, if the charged Higgs are heavier than the heavy neutrinos, then they 
 would dominantly decay into them and the corresponding charged lepton. This can lead 
 to significantly different search signal for the charged Higgs when compared to 
 conventional ones. Thus even when the charged Higgs is heavier than the top quark, 
 a presence of a light Majorana neutrino completely overwhelms the $t \,b$ decay 
 option. The charged Higgs production would be via the photon exchange mostly:
$$ p \, p \rightarrow H_1^+ \, H_1^- \to \nu_j \nu_j \ell^+_i \ell^-_i $$
where again $i=1,2,3$ while $j=i+3$. The heavy neutrino would decay via the 
off-shell charged Higgs in the 3-body decay channel $\nu_j \to \ell^{\pm} j j'$. This 
quite clearly gives a multi-lepton signature for the charged Higgs mediated by lepton-number violating interactions which again has very little or no SM background and can be 
a very unique signal of the model. For example in the case of {\bf BP1} there is a 
4-lepton channel contribution coming from the pair production of charged Higgs
\begin{align*}
 p \, p \to H_1^+ \, H_1^- \to \nu_4 \, e^+ \, \nu_4 \, e^- \to 2 e^\pm + 4 \, j + e^+ \, e^- \,\,\, .
\end{align*}
Here again this is a very interesting signal channel in the form of three 
same-sign electrons which has negligible SM background. This would prove to be an 
interesting signal \cite{Huitu:2017vye} to look for charged Higgs search in this model.

 Another unique signal of this model which differentiates it from other LR models involves the $W_R$ 
decay channels. Fig.~\ref{fig:wrbr} gives a plot of the various $W_R$ decay branching ratios 
in this model. In general LR models an important decay channel for $W_R$ boson is a $(t\,\bar{b})$ 
final state but that channel is almost absent in this case owing to the extremely small branching 
ratio as can be seen in Fig. \ref{fig:wrbr}. In fact once the heavy $T$ fermion channel opens up, a 
significant branching is into this $(T\,\bar{b})$. Thus the model opens a possibility of some 
very interesting signal topologies which are quite non-standard and can give surprisingly different 
and unique signals from production of heavy Majorana neutrino as well as charged Higgs boson 
at the LHC.

\section{Conclusion}
\label{sec:concl}
 
 In this work we have proposed a model for SM fermion mass generation through universal seesaw 
 mechanism. The model is based on a left-right symmetric framework where all the gauge symmetries 
 are spontaneously broken via $SU(2)$ scalar doublets only. The gauge symmetry is augmented with an
 additional $Z_2$ discrete symmetry which differentiates the quarks from the lepton sector. Additional heavy vector-like singlet fermions are needed for the generation of the SM quark and lepton masses through a universal seesaw mechanism. The neutrino matrix, on the other hand, can lead to two very interesting physical scenarios -- one with Majorana-like neutrinos and the other where the neutrinos are pseudo-Dirac in nature.
 
The scalar sector here may be considered as a LRS extension of lepton-specific 2HDM. The SM-like neutral Higgs boson (with mass of 125 GeV) and the lightest charged and pseudoscalar Higgs states remain light of the order of a few hundred GeV. The most stringent bounds on the charged and the pseudoscalar Higgs masses in a general 2HDM scenario come from the flavor-changing processes and two $\tau$ final state decay modes. These bound are quite relaxed in this model due to the right-handed nature of both of these light scalars and their much reduced effective couplings in this model. Hence the light charged or pseudoscalar states can easily be accommodated here which can lead to interesting collider signatures.
 
 The model also presents us with some unique collider signatures that could be observed at 
 the LHC. We consider two benchmark points in our model to highlight their signal 
 strengths. 
 In addition to interesting signal from heavy neutrino production where one gets 
 same-sign leptons and same-sign top quark pair in the same event, the model also gives 
 very unique and different 
 signal for the charged Higgs in the model. As no search has been performed at either 
 ATLAS or CMS for such event topologies, the observation of such non-standard signal 
 events at LHC could provide hints on new physics with an underlying model quite 
 different from the popular left-right models. 
 
%%%
\bigskip
\begin{acknowledgments}
%\emph {Acknowledgments:} 
This work was partially supported by funding available from the Department of Atomic Energy, 
Government of India, for the Regional Centre for Accelerator-based Particle Physics (RECAPP), 
Harish-Chandra Research Institute. AP was partially funded by the SERB National Postdoctoral 
fellowship file no. PDF/2016/000202. AP would like to thank RECAPP, Harish-Chandra Research 
Institute for their hospitality during the visit when a part of this work was done. AP would also like to 
thank Biplob Bhattacherjee for useful discussions. SKR would like to thank the CERN Theory 
Group for a short-term visit while this work was being completed and written up.
\end{acknowledgments}

%%%%%%%%%%%%%%%%%%%%%%%%%%%%%%
\end{document}